\documentclass[aps,amsfonts,amsmath,superscriptaddress,amssymb,prb,twocolumn]{revtex4}

\usepackage{graphicx}
\usepackage{dcolumn}
\usepackage{bm}
\usepackage{longtable}

\begin{document}

\title{GRADED ORBITAL OCCUPATION NEAR INTERFACES IN A $\rm La_{2}NiO_{4}-La_{2}CuO_{4}$ SUPERLATTICE}

\author{S. Smadici}
\affiliation{Frederick Seitz Materials Research Laboratory, University of Illinois, Urbana, IL 61801, USA}%

\author{J. C. T. Lee}
\affiliation{Frederick Seitz Materials Research Laboratory, University of Illinois, Urbana, IL 61801, USA}%

\author{J. Morales}
\affiliation{Frederick Seitz Materials Research Laboratory, University of Illinois, Urbana, IL 61801, USA}%

\author{G. Logvenov}
\thanks{Present address: Max-Planck-Institut, Festk$\rm \ddot{o}$rperforschung, 70569, Stuttgart, Germany.}
\affiliation{Brookhaven National Laboratory, Upton, NY 11973, USA}%

\author{O. Pelleg}
\affiliation{Brookhaven National Laboratory, Upton, NY 11973, USA}%

\author{I. Bozovic}
\affiliation{Brookhaven National Laboratory, Upton, NY 11973, USA}%

\author{Y. Zhu}
\affiliation{Brookhaven National Laboratory, Upton, NY 11973, USA}%

\author{P. Abbamonte}
\affiliation{Frederick Seitz Materials Research Laboratory, University of Illinois, Urbana, IL 61801, USA}%

\begin{abstract}
X-ray absorption spectroscopy and resonant soft x-ray reflectivity
show a non-uniform distribution of oxygen holes in a $\rm
La_{2}NiO_{4}-La_{2}CuO_{4}$ (LNO-LCO) superlattice, with excess
holes concentrated in the LNO layers. Weak ferromagnetism with $\rm
T_{c}=160~K$ suggests a coordinated tilting of $\rm NiO_6$
octahedra, similar to that of bulk LNO. Ni $d_{3z^2-r^2}$ orbitals
within the LNO layers have a spatially variable occupation. This
variation of the Ni valence near LNO-LCO interfaces is observed with
resonant soft x-ray reflectivity at the Ni and Cu L edges, at a
reflection suppressed by the symmetry of the structure, and is
possible through graded doping with holes, due to oxygen
interstitials taken up preferentially by inner LNO layers. Since the
density of oxygen atoms in the structure can be smoothly varied with
standard procedures, this orbital occupation, robust up to at least
280 K, is tunable.
\end{abstract}

\pacs{}

\maketitle

\section{Introduction}

Orbital states of transition metal (TM) oxides present the
opportunity of adjusting material properties to a specific purpose.
In practice, spin and orbital reconstructions by charge transfer
have been observed near the LCMO-YBCO
interfaces~\cite{2007Chakhalian,2006Chakhalian}, which locally have
the structure of 113 oxides between Cu and Mn atomic planes. Other
layered structures have been proposed, where strain, electronic
confinement or correlations remove the bulk orbital degeneracy
\cite{2008Chaloupka}, leaving only one orbital available for
transport, similar to the Cu $d_{x^{2}-y^2}$ orbital in
high-temperature superconductors. The $\rm LaNiO_3-LaAlO_3$
superlattices \cite{2008Chaloupka, 2009Hansmann, 2011Benckiser,
2011Boris} where the degeneracy of the $e_g$ orbitals in $\rm
LaNiO_3$ is removed, is an example of using these ideas for orbital
reconstruction and control of correlated phases of TM oxides with
the 113 structure.

It is intriguing to ask whether these ideas can be extended to 214
materials. In some bulk 113 structures (e.g. $\rm SrTiO_{3}$) the TM
ion moves away from the center of the oxygen octahedron and the
orbital degeneracy is lifted by the crystal-field splitting. Strain
can determine the relative order of thin film 113 $e_g$ orbitals if
the orbitals are degenerate in the bulk.~\cite{2008Nanda} For
instance, when the apical oxygen atoms are moved farther along the
$c$-axis, the $d_{3z^{2}-r^2}$ orbital becomes lower in energy than
the $d_{x^{2}-y^2}$ orbital. For TM ions compounds with the 214
structure, the degeneracy of $e_g$ orbitals is similarly lifted by
the crystal-field splitting. In this case, the order of $e_g$
orbitals is not modified by strain because the orbital energy change
associated with it ($\rm \sim 0.1~eV$) is relatively small compared
to the crystal-field splitting ($\rm \sim 1~eV$). In addition, in
contrast to 113 structures, additional atomic layers are present
between TM layers in 214 compounds. This raises the question of
whether 214 heterostructure interfaces will affect the occupation of
TM orbitals away from bulk properties sufficiently strongly, through
confinement effects, charge transfer or in other ways.

Here, we investigate the effect of interfaces on oxygen dopant
distribution and TM orbital occupations in a superlattice made of
$\rm La_{2}NiO_4$ (LNO) and $\rm La_{2}CuO_4$ (LCO) layers.  We
observed that oxygen atoms are doping preferentially the LNO layers,
changing the TM orbital occupation. More importantly, they can allow
a reflection suppressed by symmetry, with the dopant-induced
scattering separated in momentum from other charge scattering. A
spatially variable Ni valence and $d$-orbital occupation within the
LNO layers is observed in this way with resonant soft x-ray
scattering. We therefore found that superlattice interfaces can
significantly affect the occupation of the TM orbitals in 214
oxides.

This corroborates the observation that oxygen content can determine
the transport properties of 113 TM oxide heterostructures. For
instance, $\rm LaAlO_3$ films on $\rm SrTiO_{3}$ (STO) substrates
are insulating or metallic, depending on the number of oxygen
vacancies.~\cite{2007Herranz} Our results show that, to understand
the properties of oxide heterostructures, it might be necessary to
consider the oxygen dopant distribution, not only the average
content. Since the oxygen interstitial distribution is relatively
easily modified with standard procedures, this orbital occupation in
214 LNO-LCO superlattices is tunable.

\section{Experiments}

\subsection{Superlattice structure}

The superlattice, made of LNO and LCO layers on a STO substrate, was
grown by molecular beam epitaxy at Brookhaven National Laboratory.
Atomic force microscopy (AFM) measurements on a Dimension 3100
instrument showed good superlattice surface quality with an average
rms roughness over a $\rm5~\mu m \times 5~\mu m$ area of $\rm
0.53~nm$. The sample has La in both layers; therefore it is not
possible to use scattering near the La edge to characterize the
interface roughness.~\cite{2009S}

Off-specular hard x-ray scattering measurements (not shown)
confirmed that the superlattice is fully strained in the $a-b$ plane
to the STO substrate with $a_{s}=b_{s}=\rm3.915\pm0.035~\AA$. A
comparison to measurements on bulk LNO and LCO single crystals
\cite{1994Tranquada,1987Shirane} indicates that the LNO layers are
under tensile stresses of 1.5 $\%$ and 0.6 $\%$ along the $a$ and
$b$ directions of the STO substrate, while the LCO layers are under
tensile stresses of 3.4 $\%$ and 2.7 $\%$ along the $a$ and $b$
directions.

\begin{figure}
\centering\rotatebox{0}{\includegraphics[scale=0.65]{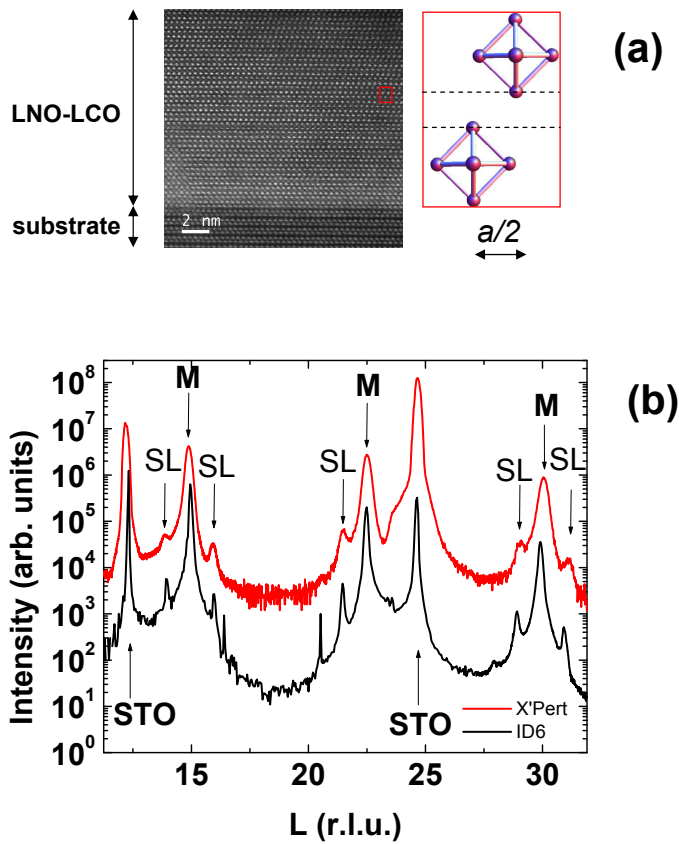}}
\caption{\label{fig:Figure1} (color online) (a) STEM high angle
annular dark field image of a LNO-LCO superlattice showing epitaxial
growth. The right panel expands the image area enclosed by the
rectangle. The bright features in the image are La atoms in the LaO
planes, represented by dashed lines between the $\rm TM-O_6$
octahedra. The oxygen atoms are not visible. The experimental
conditions were $\rm 1.3~\AA$ probe size at 200 kV, $\rm 28~mrad$
convergence angle, and $\rm 64-341~mrad$ collection angle. (b) Wide
angle hard x-ray measurements made at Advanced Photon Source at
Argonne National Laboratory (7.5 keV) and with an X' Pert
diffractometer (8.05 keV) showing superlattice peaks (SL) near the
main reflections (M) indexed with respect to the measured
superperiod. Substrate peaks (STO) are also visible.}
\end{figure}

Hard x-ray reflectivity (HXR) scans (Fig. 1b), made at beamline ID6
at the Advanced Photon Source at Argonne National Laboratory and
using a laboratory X'Pert x-ray source, showed pronounced
superlattice peaks. The average monolayer thickness was $\rm
6.470\pm0.035~\AA$, defined as 1 ``molecular layer" (ML), and
corresponding to half the average of LNO and LCO unit cells $c$-axis
parameters. The number of ML in a superperiod was determined by
measuring the ratio of momentum transfers for superlattice
reflections satellites (SL) to momentum transfers of Bragg peaks (M
in Fig. 1b) corresponding to 1 ML. The average of ID6 and X'Pert
measurements showed a superlattice period of (7.45 $\pm$ 0.08) ML
and (7.42 $\pm$ 0.14) ML, respectively, indicating an approximately
equal mixture of 7 ML and 8 ML superperiods.

The average ML thickness, number of MLs in a superperiod and number
of superperiods (12 deposition repeats), give a total superlattice
thickness of $\rm 577 \pm 10~\AA$, consistent with the result
obtained from thickness oscillations (Fig. 2) of $\rm 570 \pm
20~\AA$.

The superlattice deposition sequence had LNO and LCO layers of
nominally equal thicknesses. Indeed, HXR measurements showed a
step-like feature at $L=2$, with the scattering momentum $Q = 2 \pi
L/ c$ indexed with respect to the superlattice superperiod $c$ for
low-$Q$ range, on a large sloping background (ID6 curve in Fig. 2).
This arises from the slightly different $c$-axis parameters of LNO
and LCO unit cells. Scattering at the $L=2$ superlattice reflection
is suppressed for equal LNO and LCO layer thicknesses. Thus, the
thicknesses of the LNO and LCO layers will be considered equal in
the following analysis. However, the $L=2$ peak, suppressed for
non-resonant HXR scattering, is observed for resonant x-ray
scattering at the Ni edge, but not at the Cu edge. This,
\emph{regardless of any irregularities of the structure}, highlights
a difference in the initial and intermediate state distributions of
resonant scattering at the Ni edge, that is a difference between the
distributions of atomic cores and Ni valence electronic states. In
Section II.E, it will be shown how this can happen at the Ni edge
(Fig. 2).

\begin{figure}
\centering\rotatebox{270}{\includegraphics[scale=0.5]{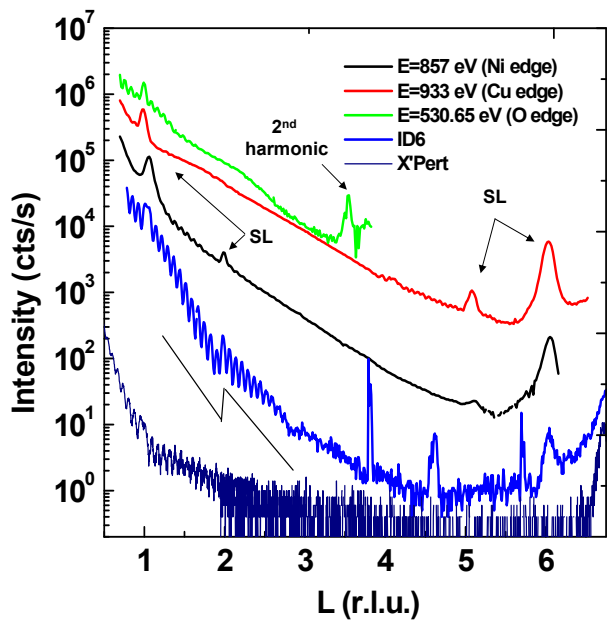}}
\caption{\label{fig:Figure2} (color online) $L$ scans of the low
momentum range for hard and soft x-ray scattering. HXR measurements
show strong thickness oscillations, with a peak at $L=1$ and a step
at $L=2$ (black line) for ID6 data. The peaks at L=3.8, 4.6, and 5.7
are stray reflections. Soft x-ray scattering shows pronounced
superlattice (SL) peaks for $L=1$ at O K, Ni L and Cu L edges. The
Ni $L=1$ peak is shifted because of refraction. The $L$=2 reflection
is much larger at the Ni edge compared to the Cu edge. This
observation is confirmed by detailed energy-resolved measurements
(Fig. 6) which are discussed below. In contrast, the absence of the
$L$=2 reflection in measurements at the oxygen edge is caused by the
energy used (see Fig. 5 for a complete measurement). Smaller
differences can be observed between the measurements at the Cu and
Ni edges, for instance a slight indication of a peak at $L=4$ at the
Cu edge. The small difference in intensity at $L=4$ can be explained
using the different polarizations of Ni and Cu $d$-orbitals, unlike
the opposite and larger difference at $L=2$ (Sec. II E). The
intensity of SL reflections at $L=3$ and $L=4$ is small, consistent
with the calculated dependence of SL structure factors amplitude on
$L$. The $L=5$ and $L=6$ SL peaks are larger because of the
proximity to the Bragg peak corresponding to 1 ML (off $x$-axis in
this figure).}
\end{figure}

\subsection{Superlattice magnetism}

Undoped bulk LNO and LCO are antiferromagnetic insulators with two
and one unoccupied TM orbitals and nominal $S=1$ and $S=1/2$ spins
on the $\rm Ni^{2+}$ and $\rm Cu^{2+}$ ions, respectively. Bulk $\rm
La_{2}NiO_{4.03}$ shows a weak ferromagnetism below $\rm 65~K$ with
a net magnetic moment oriented along the $c$-axis of approximately
$\rm 0.04 ~\mu_{B}/Ni~ion$ at $\rm 4.2~K$.~\cite{1996Demourgues} The
ferromagnetism is not the consequence of oxygen doping; the $\rm
NiO_6$ octahedra in undoped bulk LNO tilt in the low-temperature
tetragonal phase, misaligning the spins and making the material
weakly ferromagnetic.~\cite{1992Yamada} In doped LNO the
ferromagnetic phase appears below $\rm \sim
150-200~K$.~\cite{1992Yamada}

SQUID measurements of the superlattice, done with a Quantum Design
instrument, showed a ferromagnetic moment below $\rm T=160~K$ (Fig.
3), with $\rm \lesssim 0.05~\mu _{B} / (Ni~ion)$, consistent with
the magnitude of bulk LNO ferromagnetism. The SQUID measurements
also did not show a superconducting anomaly in the magnetic moment
down to $\rm 10~K$, which sets an upper limit on hole doping of the
LCO layers of 0.06. \cite{2002Fujita}

The consistency of the magnitude of the moment and temperature of
the transition with that of bulk LNO suggest that the superlattice
ferromagnetism is likely confined to the LNO layers, where the
magnetic moment has a similar origin to that of bulk LNO. The
necessary corresponding modifications to the magnitude of Ni spins
near the interfaces from different Ni orbital occupations, discussed
later and shown schematically in Fig. 8b, do not appear to result in
large additional ferromagnetic moments. In this case, this is likely
because exchange correlations decay spatially very fast and the net
magnetic moment of $\rm NiO_2$ and $\rm CuO_2$ layers in bulk
compounds is small or zero, respectively. The temperature variation
of the SL ferromagnetism will invalidate a candidate explanation for
the difference at L=2 in the resonant soft x-ray scattering at the
Ni and the Cu edges (Sec. II E). We will use the smallness of the
superlattice ferromagnetic moment in Section III.A.

\subsection{Hole distribution}

Oxygen doping bulk LNO and LCO places oxygen interstitials between
the LaO layers of the structure.~\cite{1989Jorgensen} The electronic
structures and orbital orientations of bulk LNO and LCO, studied
with x-ray absorption spectroscopy (XAS) experiments on bulk
nickelates \cite{1996Pellegrin, 1991Kuiper,1992Yamada} and cuprates
\cite{1991Chen}, showed doping-dependent upper Hubbard bands (UHB)
and mobile carrier peaks (MCP).

XAS measurements showed a non-uniform distribution of holes in the
SL and the favorable conditions for a large Ni valence modulation.
We used XAS in total fluorescence yield (TFY) and total electron
yield (TEY) modes to characterize the hole distribution in the
LNO-LCO superlattice (Fig. 4a). Measurements were made at beamline
X1B at the National Synchrotron Light Source at Brookhaven National
Laboratory. To determine which peak belongs to which layer, the
short probing depth of TEY compared to TFY was used to isolate the
signal from the shallower layer of the superperiod, which is known
from the deposition sequence to be LNO. For energies in the range of
a few hundred eV, as in our case, the TEY probing depth is less than
$\rm 100~\AA$. In contrast, the penetration depth of TFY at these
energies is thousands of $\rm \AA$, effectively probing the entire
superlattice. Therefore, the peaks that are present in TFY but not
TEY originate from the deeper layer of the superperiod pair, which
is the LCO layer.

Fig. 4a shows TFY and TEY measurements on the superlattice at room
temperature. The lowest energy peak that originates in the LCO
layers aligns with the upper-Hubbard band (UHB) peak of bulk
LCO,~\cite{1991Chen} while the lowest-energy peak in the LNO layers
aligns with the mobile carrier peak (MCP) of bulk
LNO.~\cite{1996Pellegrin,1991Kuiper} This suggests that the LCO
layers of the superlattice are undoped, consistent with the
magnetization measurements in Section II.B, while the LNO layers are
doped by excess oxygen. The LCO MCP peak begins to be clearly
present at $x \rm \sim 0.03$ in bulk $\rm
La_{2-x}Sr_{x}CuO_4$~\cite{1991Chen}; since no such peak is visible
for the LNO-LCO superlattice, it must be that $x_{LCO} \rm < 0.03$.
Similarly, the LNO UHB is present up to $x\sim 0.2$ in bulk $\rm
La_{2-x}Sr_{x}NiO_4$~\cite{1996Pellegrin}; since no peak is seen in
TFY or scattering other than LCO UHB in this energy range, $x_{LNO}
\rm > 0.2$ or oxygen doping $\delta > 10 $ $\%$. We conclude that
the doped holes are localized in the LNO layers of the superlattice.

Even with a large number of holes in the LNO layers, the
superlattice remains insulating (Fig. 3). Assuming that holes are
shared equally between Ni and O atoms, $x_{LNO}\rm~> 0.2$ implies
that the average Ni valence increases from 2+ in undoped LNO to
larger than $2.1+$ in oxygen-doped LNO. It can, in principle,
accommodate a substantial modulation of the Ni valence within the
LNO layers. This modulation is observed by scattering at the Ni edge
(Section II.E).

\subsection{Scattering at the oxygen edge}

Resonant soft x-ray scattering (RSXS) can probe bulk charge
order.~\cite{2006Abbamonte,2005Abbamonte,2004Abbamonte,2002Abbamonte}
Its energy and momentum selectivity makes it a useful technique for
also probing the spatial distributions of orbitals of different
energies in nanometer-sized structures. Measurements on the LNO-LCO
superlattice at the oxygen K, Ni and Cu L edges (see Fig. 2 for
selected resonant energies) were made at beamline X1B at the
National Synchrotron Light Source in an ultra-high vacuum
diffractometer with a $\pi$-polarized incident beam. Scattering
measurements were made at 90 K, unless specified otherwise. The beam
was focused to an approximately $\rm 1~mm \times 1~mm$ spot, which
is much smaller than the sample size.

RSXS at the oxygen K edge probes the spatial distribution of the
occupation of oxygen $p$-states and illustrates one reason for the L
dependence of resonant intensities of SL reflections. Measurements
were made in specular geometry with a variable energy $E$ and
momentum perpendicular to the surface $Q = 2 \pi L/ c$, with
reflections indexed with respect to the superperiod $c$ for low-$Q$
range (for other examples of RSXS on superlattices, see
~\cite{2007S,2009S}). Figs. 4b and 5 show reflectivity measurements
at the oxygen edge for L=1 and L=2, respectively. Refraction effects
and small superlattice imperfections slightly shift the peaks from
integer values. Peaks at 529 eV and 531.5 eV are visible for the
momentum held fixed at L=1 (Fig. 4b), indicating a spatial
modulation of the same LNO MCP and LCO UHB states that are visible
at 528.5 eV and 530.5 eV in FY measurements (Fig. 4a). A third peak
is present for L=2 at 533.5 eV (Fig. 5); this peak will be addressed
in Section III.B. The ratio of the scattering intensities for LNO
MCP and LCO UHB energies is different for L=1 and L=2. Specifically,
$\rm I_{LNO,MCP}/I_{LCO,UHB}=1.1$ for L=1 and 0.4 for L=2.

\begin{figure}
\centering\rotatebox{270}{\includegraphics[scale=0.35]{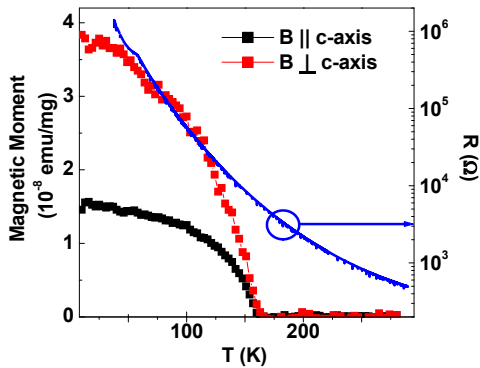}}
\caption{\label{fig:Figure3} (color online) Magnetic moment
measurements show a ferromagnetic transition at $\rm 160~K$. A
diamagnetic contribution has been subtracted from the data. The
applied magnetic field was 100 Oe. Resistivity measurements show an
insulating behavior down to $\rm 38~K$ with a small anomaly in slope
at $\rm 58~K$.}
\end{figure}

The intensity of a superlattice reflection in the single-scattering
approximation is given by $I=A|\hat{\epsilon}_{f}^{*}S_{L}
\hat{\epsilon}_{i}|^2$, where
\begin{eqnarray}
S^{ij}_{L}=\sum_{l,n} \langle t_{l}^{n}\rangle
f^{ij}_{n}(\omega)e^{i2 \pi L z_{l}/c} \label{eqn:Equation1}
\end{eqnarray}
is the structure factor, $\langle t_{l}^{n}\rangle$ is the average
over a layer of the occupation factor $t_{l}^{n}$ (equal to $0$ or
$1$), which identifies the valence and element $n$ to use in Eq. 1
for each site in layer $z_l$ (e.g. $\rm Ni^{2+}$ or $\rm Ni^{3+}$),
and $f^{ij}_{n}(\omega)$ is the atomic form factor for valence and
element $n$, with the indices $i$ and $j$ accounting for X-ray
birefringence. Defining the variables $\{t_{l}^{n}\}$ in Eq. 1
separates the spatial ($l$) and energy ($\omega$) dependence of the
scattering factor. This is justified by the very weak dependence of
atomic form factors on momentum for soft x-ray energies. In
comparing the intensities of LNO MCP and LCO UHB features, we
neglect self-absorption effects and the change with geometry of $A$
(e.g., footprint effects) since the scattering geometries and
energies are very close. The variation with $L$ of the intensity
ratio $\rm I_{LNO,MCP}/I_{LCO,UHB}$ can, in principle, be due to
either the polarization dependence of the form factors $f^{ij}_n$ or
the spatial distribution of the occupation factors $t_{l}^{n}$.

Bulk LCO has hole orbitals oriented in the $a-b$ plane
\cite{1991Chen}, while the holes in bulk LNO have a substantial
component oriented along the
$c$-axis~\cite{1991Kuiper,1995Kuiper,1998Kuiper}. The variation of
the intensity ratio $\rm I_{LNO,MCP}/I_{LCO,UHB}$ with $L$ can be
qualitatively explained by assuming a polarization of hole orbitals
in the superlattice LCO and LNO layers that resembles that of bulk
LCO and LNO. Specifically, the scattering intensity of a
horizontally-polarized beam off intermediate states oriented along
$a-b$ and $c$-axis is proportional to $\rm sin^{4}(\theta)$ and $\rm
cos^{4}(\theta)$ respectively, where $\theta$ is the angle between
the incident beam direction and the sample surface. The intensity
for the LCO state will be proportional to $\rm sin^{4}(\theta)$. In
contrast, for the LNO intermediate state, with a substantial
component oriented along $c$-axis, the scattering intensity has a
term proportional to $\rm cos^{4}(\theta)$ and can be expected to
increase slower or even decrease as $\theta$, and therefore $L$,
increases. Then, to explain the variation with $L$ of $\rm
I_{LNO,MCP}/I_{LCO,UHB}$ observed in the measurements, it is
sufficient to assume, based on the relatively small influence of
strain in 214 structures (Section I), that the hole $p$-orbital
orientations in LNO and LCO layers are similar to those in bulk
crystals.

\begin{figure}
\centering\rotatebox{0}{\includegraphics[scale=0.65]{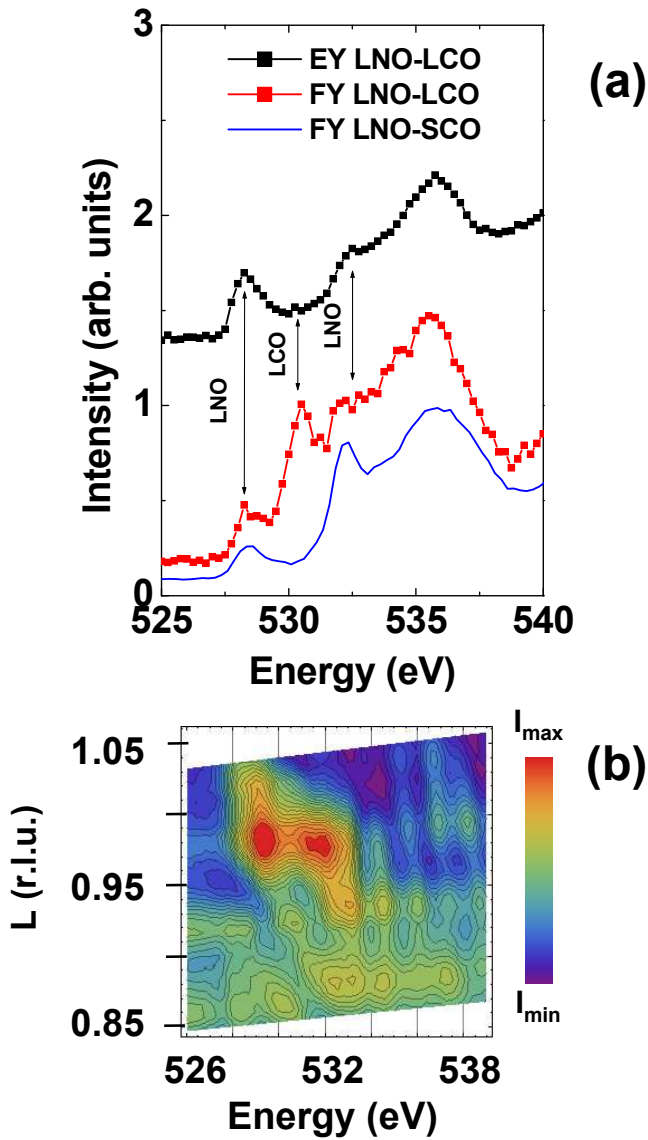}}
\caption{\label{fig:Figure4} (color online) (a) Electron and
fluorescence yield and the correspondence of the peaks to the
superlattice layers. The LCO UHB peak disappeared completely for an
LNO-SCO superlattice grown on an STO substrate, showing that there
is no STO substrate or LNO layer contribution at the LCO UHB energy.
From detailed measurements (not shown) on LNO-SCO superlattices LNO
MCP and LCO MCP have different energies. (b) RSXS intensity
(logarithmic scale) at 90 K at the O edge shows two peaks
corresponding to the first two peaks in TFY, with the lower (529 eV)
and higher (531.5 eV) energy peaks from the LNO and LCO layers,
respectively. The different $L$ profiles are due to an interference
effect and different layer $c$-axis parameters.}
\end{figure}

The oxygen $p$ orbitals hybridize with the TM orbitals inside the
$\rm TM-O_{6}$ octahedra (Fig. 1a). Therefore, the polarization of
the oxygen form factors $f^{ij}$ is related to that of TM form
factors and a similar variation of the scattering intensity with $L$
at the TM edges is expected. However, in contrast to the case of
oxygen edge LNO MCP and LCO UHB states, in the next section it will
be shown that the evolution with $L$ of the scattering intensities
for Ni and Cu orbitals cannot be explained with polarization of form
factors alone.

\subsection{Scattering at transition metal edges}

Resonant scattering at TM edges probes the $d$-states of Cu and Ni
ions. $L$ scans on resonance at these energies (Fig. 2) show the
expected peaks at $L$=1. The $L$=2 peak is not visible at the Cu L
edge in Fig. 2; however, the $L$=2 peak is unexpectedly pronounced
at the Ni edge. Detailed energy plots of this momentum region, shown
in Fig. 6, confirm the large difference $\rm
I(L=2)/I(L=1)|_{Ni}>>I(L=2)/I(L=1)|_{Cu}$ between the L=2
reflections at the two edges: the peak at the Ni edge is 100 $\%$
more intense than the background, compared to only 6 $\%$ at the Cu
edge. This superlattice reflection was observed for sample azimuthal
orientations with incident beam polarization at $\rm 45^{\circ}$ and
along the tetragonal $a-b$ axes (Fig. 7a). In contrast to the
temperature dependence of the magnetic moment in Fig. 3, the
intensity of the $L$=2 peak at the Ni edge does not change with
temperature between $\rm 90~K$ and $\rm 280~K$ (Fig. 7b). Therefore,
this peak is not related to ferromagnetism.

Contrast the observation of the larger $L$=2 intensity at the Ni
edge compared to the Cu edge with the \emph{opposite} trend at the O
edge for LNO MCP and LCO UHB states, in which the LNO state
intensity is suppressed at larger $L$. Therefore, the difference in
the intensity ratio $\rm I(L=2)/I(L=1)|_{Ni,Cu}$ for scattering at
Ni and Cu edges cannot be explained using the difference in the
orientation of $p$-states in the LNO and LCO layers, noted at the
oxygen edge and a similar polarization of $d$-state TM orbitals
hybridized with oxygen orbitals in their respective layers. The
polarization of the form factor $f^{ij}_{n}(\omega)$ cannot be the
explanation; instead we must look to a difference in the occupation
factors $t_{l}^{n}$ for Ni and Cu $d$-orbitals.

This difference originates in the different dependence of the Ni and
Cu valence on doping. LNO resembles a Mott-Hubbard insulator more
than LCO.~\cite{1985Zaanen} This is visible in the small degree of
contrast in TFY at the Cu edge with doping of bulk
LCO~\cite{1992Chen} and the large contrast at the Ni edge with
doping bulk LNO~\cite{1998Kuiper}. Therefore, the Ni $d$-orbitals
participate in the hole doping process more than the Cu
$d$-orbitals. The Cu $d$-states occupation factor $t_{l}^n$ follows
the distribution of the Cu ions because their valence remains close
to $\rm Cu^{2+}$. In contrast, the substantial oxygen doping in the
LNO layers (Section II.C) and the strong coupling between hole
doping and Ni $d$-states, empties additional Ni orbitals, thus
raising the Ni ion valence toward $\rm Ni^{3+}$.

Therefore, unlike the Cu case where oxygen doping is absent, the
spatial distributions of both Ni atoms \emph{and} oxygen
interstitials determine the occupation of Ni $d$-states that is, the
magnitude of $t_{l}^n$. If the oxygen dopants are not uniformly
distributed within the LNO layers, a variable Ni valence results,
i.e. spatial distributions $t_{l}^n$ for the doped orbitals
different from those of the Ni ions. In this case, as pointed out in
Section II.A, since the intermediate $d$-states do not follow the
distribution of the underlying structure represented by the core
atoms, the scattering at $L$=2 is not suppressed. In the next
section, we will use this observation and Eq. 1 to determine the
orbital occupancies $t_{l}^n$ in the LNO layers.

\section{Discussion}

Our main finding does not depend on structural irregularities. The
distributions of Cu and Ni atoms, and therefore of core $p$-states,
are complementary (the layer occupations are $\langle
t_{l}^{Cu}\rangle+\langle t_{l}^{Ni}\rangle=1$, for each layer $l$,
where $\langle t\rangle$ is the occupation factor from Eq. 1), as
supported by the observation of a crystalline SL structure with Cu
and Ni atoms occupying their normal sites within the 214 unit cell.
Complementary distributions have the same absolute value of the
Fourier transform amplitude at L=2. Therefore, if one does not
consider differences in the Cu and Ni $d$-state occupations, since
the core $p$-states (the initial states of the RSXS transition)
distributions are complementary, for any value of superlattice
superperiod, specific model of the distribution of interface atoms
(not necessarily Gaussian as we assume in Sec. III B) or size of
interface roughness, the L=2 peak cannot be suppressed at the Cu
edge without being also suppressed at the Ni edge. To explain the
difference in scattering at L=2, it is necessary to consider
differences in the $d$-states (the intermediate states of the RSXS
transition) that is, in the electronic properties of the LCO and LNO
layers.

\begin{figure}
\centering\rotatebox{270}{\includegraphics[scale=0.7]{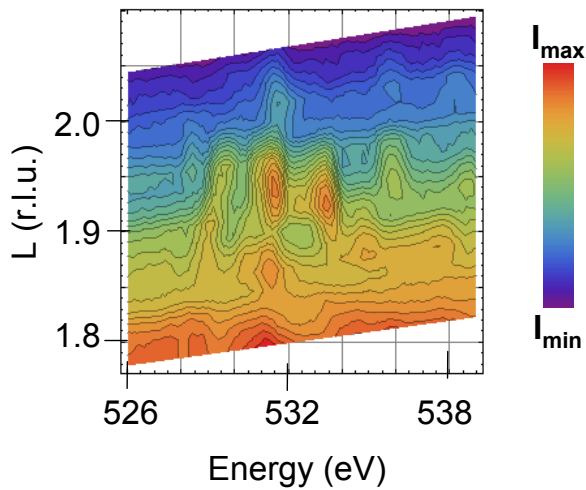}}
\caption{\label{fig:Figure5} (color online) Intensity of the $L$=2
reflection at the O edge at 90 K (logarithmic scale). The three
peaks are (in order of increasing energy): LNO holes at E=529.5 eV,
LCO UHB holes at E=531.5 eV and a feature at E=533.5 eV, discussed
briefly in Section III.B. Unlike the scattering at $L$=1, the LNO
MCP peak has substantially smaller intensity than the LCO UHB peak.}
\end{figure}

The difference in the L=2 scattering at the Cu and Ni edges cannot
be attributed to a different resonant form factor $f^{ij}(\omega)$
for the Cu and Ni edges. The energy $\omega$ and momentum $L$
variables are separable in the soft X-ray scattering cross-section
(Eq. 1); therefore, since Cu and Ni atoms scatter at L=1,
considering only the different form factors $f^{ij}(\omega)$ will
imply that both scatter at L=2, contrary to observations.

The conclusion that the Ni valence is not constant in the LNO layers
is obtained from comparing the presence of L=2 scattering at the Ni
edge with its absence at the Cu edge in the scattering model of Sec.
III A. The measured ratio I(L=2)/I(L=1) at the Ni edge and use of
specific SL structure models are necessary for a quantitative
estimate (Sec. III B).

\subsection{Scattering model}

\begin{figure}
\centering\rotatebox{0}{\includegraphics[scale=0.7]{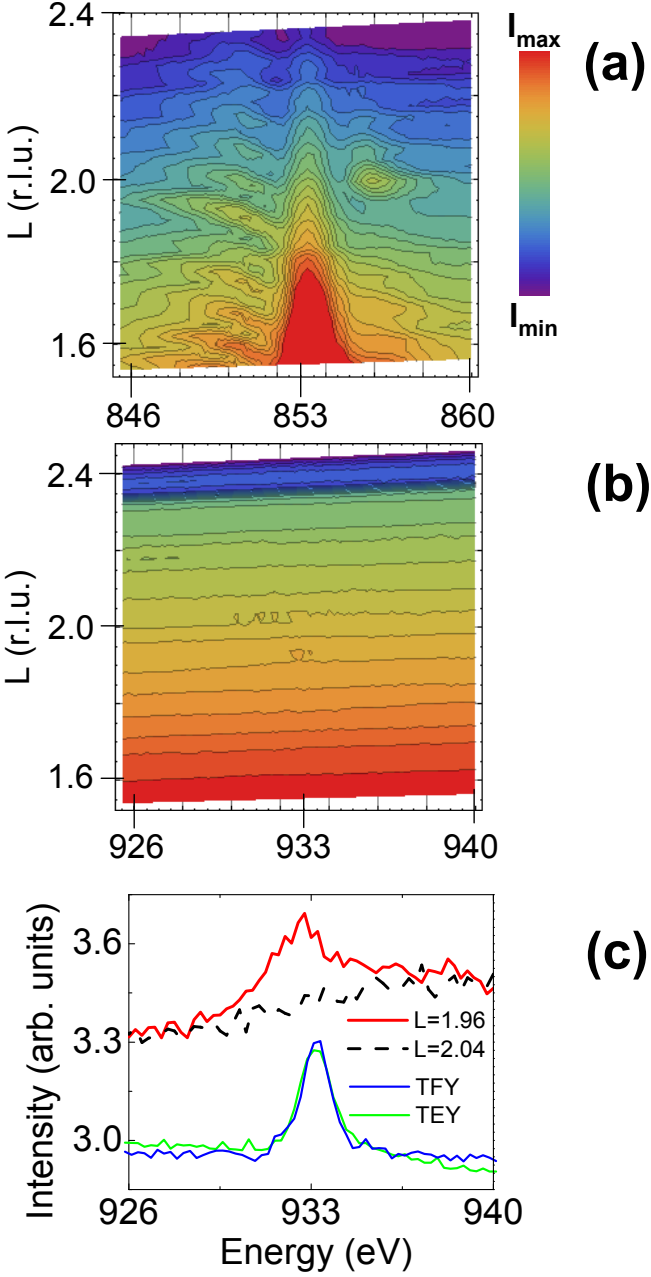}}
\caption{\label{fig:Figure6} (color online) (a) Scattering intensity
(logarithmic scale) at La $\rm M_4$ (E=853.25 eV) and Ni L$\rm _3$
(E=856 eV) edges at $L$=2. The increase in the intensity over
background at the Ni edge is 100 $\%$. The features below 853 eV are
thickness oscillations. No off-resonant intensity tail is observed
near $L=2$ at the Ni edge, showing that this reflection originates
in an electronic state and not from, for instance, selective removal
of Ni ions from the LNO layers. (b) Scattering intensity
(logarithmic scale) over a wide range at the Cu edge near $L=2$ for
comparison. The energy range (14 eV) is chosen the same as in (a).
(c) The area of interest at the Cu edge magnified, compared to TFY
and TEY. A very small peak is visible for $L$=1.96, with a relative
change in intensity over background of only 6 $\%$. Because of the
large sloping background an $L$ scan is less useful in showing this
small peak.}
\end{figure}

Calculating the energy dependence of Ni edge scattering requires a
knowledge of the energy and doping dependence of the scattering
factors $f^{ij,\rm{Ni}}_{l}(\omega)$ in the LNO crystalline lattice.
Band structure calculations of the Ni density of states variation
with doping with LSDA+U models~\cite{2007Yamamoto} cannot reproduce
the XAS experiments in the low-doping range~\cite{1998Kuiper}. A
detailed cluster analysis of energy levels of $\rm Ni^{2+}$ and $\rm
Ni^{3+}$ ions in an octahedral crystal field shows that the
degeneracy of the x-ray transitions is lifted, which results in a
very complex transition spectrum.\cite{1990deGroot} This is visible
in Fig. 8a (top) for $\rm Ni^{2+}$ and $\rm Ni^{3+}$ ions in a cubic
field. The scattering at $L$=2 at the Ni $\rm L_3$ and $\rm L_2$
edges is nevertheless made of only one peak (Figs. 7c inset and 8a).
This suggests that a simplified scattering model is warranted.

The scattering at the Ni edge can be analyzed using Eq. 1 and a
simplified model. We define the functions $f^{ij}_{n}(\omega)$
first. The XAS and $L=2$ scattering energy profile (Fig. 8a) do not
resemble the density of states of the intermediate unoccupied Ni
$d$-states; for instance, each of the two peaks in the XAS at the Ni
L$\rm_3$ edge, following immediately after the La edge at $\rm
853.5~eV$ in Fig. 8a, has both $x^{2}-y^2$ and $3z^{2}-r^2$
character.~\cite{1998Kuiper} However, the state into which the
electron is excited in the transition is well-defined and will be
used as a label. Therefore, the transitions at the Ni L$\rm_3$ edge
are labeled by the intermediate $d$-states involved in the 2-step
RSXS process and organized into four main groups based on the spin
of this intermediate state (up or down) and its orbital symmetry
($d_{3z^{2}-r^2}$ and $d_{x^{2}-y^2}$). For undoped LNO (or a $\rm
Ni^{2+}$ valence), two of these states are unoccupied
($d_{3z^{2}-r^2,\uparrow}$ and $d_{x^{2}-y^{2},\uparrow}$). Doping
toward a $\rm Ni^{3+}$ valence is a complex process that cannot be
represented as sequential emptying of additional orbital levels.
However, because of the isotropic nature of the holes in bulk
LNO~\cite{1991Kuiper,1995Kuiper}, it is expected that both the other
two $d$-states ($d_{3z^{2}-r^2,\downarrow}$ and
$d_{x^{2}-y^{2},\downarrow}$) will be emptied. Indeed,
polarization-dependent measurements of bulk LNO at the Ni edge show
that these states are approximately equally emptied for
$0<x<0.12$~\cite{1998Kuiper}, becoming available for the scattering
process. As at the oxygen edge, it is assumed that the bulk $f^{ij}$
can be used for the layers of the LNO-LCO superlattice since, as
pointed out in Section I, strain will not change the relative order
of $d$-orbital energies in compounds with the 214 structure.
Therefore, the atomic form factors for Ni ions in the layers are:
\begin{eqnarray}
f^{ii}_{2+}(\omega)=f^{ii}_{d_{3z^{2}-r^{2}},\uparrow}+f^{ii}_{d_{x^{2}-y^{2}}\uparrow},
\label{eqn:Equation3}
\end{eqnarray}
\noindent and
\begin{eqnarray}
f^{ii}_{(2+) \rightarrow
(3+)}=\frac{f^{ii}_{d_{3z^{2}-r^{2}},\downarrow}+
f^{ii}_{d_{x^{2}-y^{2}},\downarrow}}{2}, \label{eqn:Equation4}
\end{eqnarray}
\noindent where $f^{ij}_{2+}(\omega)$ is the form factor for $\rm
Ni^{2+}$ ions, $f^{ij}_{(2+) \rightarrow (3+)}(\omega)$ is the term
added with increasing the Ni valence to $\rm Ni^{3+}$, and the
off-diagonal terms have been neglected because the ferromagnetic
moment is small (Section II.B). These equations reflect the
observation that in undoped LNO the $d_{3z^{2}-r^2},\uparrow$ and
$d_{x^{2}-y^2},\uparrow$ levels are empty and that doping empties in
approximately equal amounts the orbitals of the opposite spin.

\begin{figure}
\centering\rotatebox{0}{\includegraphics[scale=0.65]{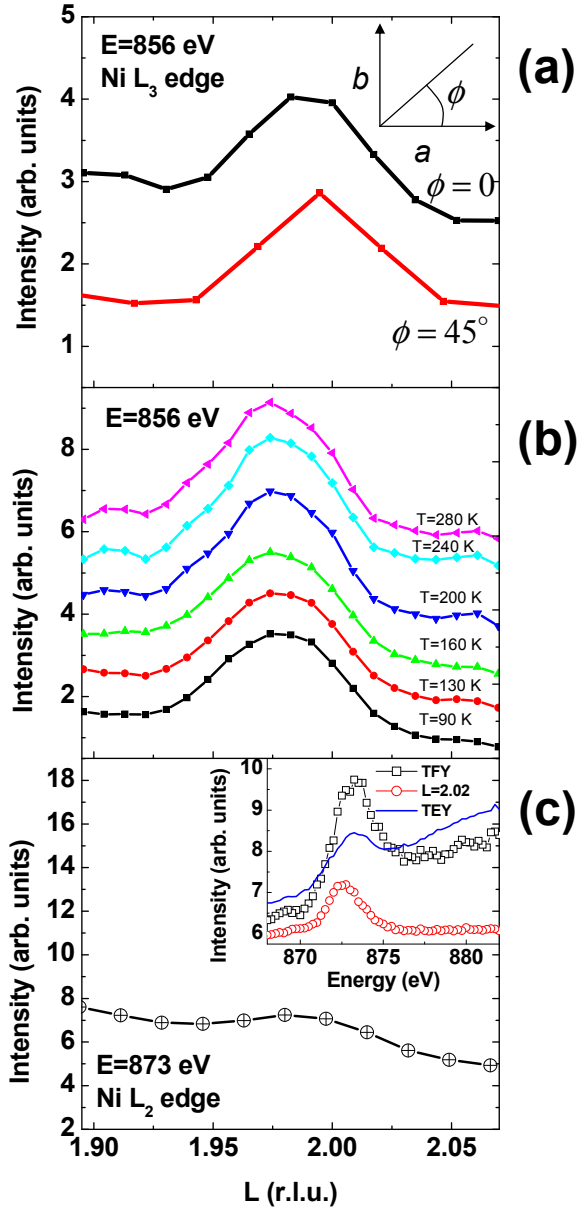}}
\caption{\label{fig:Figure7} (color online) (a) $L$ scans at the Ni
$\rm L_3$ edge show the $L=2$ peak for sample azimuthal orientations
with incident beam polarization at $\rm 45^{\circ}$ and along the
tetragonal $a-b$ axes. (b) The $L=2$ peak at the Ni $\rm L_3$ edge
persists up to at least 280 K. As will become clear in Section
III.A, this is due to the larger energy necessary to move the oxygen
dopants in the structure compared to the thermal energy needed to
destroy a spin order. (c) The $L=2$ peak at the Ni $\rm L_2$ edge.
The inset shows the energy profile of this peak (circles) compared
to TFY (squares) and TEY (blue line).}
\end{figure}

The other term in Eq. 1 is $\langle t^{n}_{l}\rangle$. To model the
momentum dependence of scattering through the orbital occupancies
$t_{l}^n$, we introduce the variables $p^{3+}_l$ and
$p^{2+}_l=s_{l}-p^{3+}_l$, representing the layer-resolved average
number of Ni atoms with nominal $\rm Ni^{3+}$ and $\rm Ni^{2+}$
valences in 1 ML over a $a_{s} \times b_{s}$ area, where $s_{l}$ is
the distribution of Ni ion cores. For instance, the probability that
the Ni ions have the nominal $\rm Ni^{3+}$ valence is represented by
$0 \leq p^{3+}_l \leq s_{l}$. From Eq. 3, the
$d_{3z^{2}-r^2},\downarrow$ and $d_{x^{2}-y^2},\downarrow$ levels
are unoccupied with a distribution following that of $p^{3+}_l$.
Thus, the average over each layer of the occupation factors
$t_l^{n}$ for Ni ions are $\langle
t^{d_{3z^{2}-r^{2},\uparrow}}_{l}\rangle=s_{l}$, $\langle
t^{d_{x^{2}-y^{2},\uparrow}}_{l}\rangle=s_{l}$, $\langle
t^{d_{3z^{2}-r^{2},\downarrow}}_{l}\rangle=p^{3+}_{l}$ and $\langle
t^{d_{x^{2}-y^{2},\downarrow}}_{l}\rangle=p^{3+}_{l}$.

Then, the average Ni scattering factor in layer $l$ depends on the
variable number of unoccupied orbitals as:
\begin{align}
f^{ij,\rm {Ni}}_{l}(\omega)\equiv \langle t_{l}^{n}\rangle
f^{ij}_{n}\Big|_{n=\rm{Ni}}=s_{l}f^{ij}_{2+}(\omega)+
p^{3+}_{l}f^{ij}_{(2+) \rightarrow
(3+)}(\omega),\label{eqn:Equation2}
\end{align}
\noindent where $f^{ij}_{2+}(\omega)$ has the distribution of Ni
cores $\{s_l\}$, to which spatially-dependent doping adds an
additional term proportional to $p^{3+}_{l}$ and $f^{ij}_{(2+)
\rightarrow (3+)}(\omega)$. Therefore, the spatial symmetry of
$f^{ij,\rm{Ni}}_{l}(\omega)$ is determined not only by the
distribution of Ni core states $\{s_l\}$, but also by the
intermediate $d$-state through $p^{3+}_l$, the dopant distribution.
The symmetry requiring the $L$=2 scattering to be small, noted in
Sections II.A and II.E, is lifted in this way, by making the
valences, and therefore the scattering factors, of Ni ions in outer
and inner LNO layers different (Fig. 8b). This allows the $L$=2
reflection at the Ni edge. In contrast, because there is no change
in the valence of the Cu ions (Section II.E), the Cu scattering
factor $f^{ij,\rm {Cu}}_{l}(\omega)$ depends on structure only,
through $\{1- s_{l} \}$, the complementary distribution to $\{ s_{l}
\}$, and the $L=2$ reflection remains suppressed.

\subsection{Ni valence modulation}

We use this model to arrive at an estimate of the magnitude of Ni
valence modulation in the LNO layers. Self-absorption is neglected
because the calculated attenuation lengths are $\rm 1400~\AA$ and
$\rm 1600~\AA$ at the Ni and Cu edges, respectively, compared to the
total SL thickness of $\rm 575~\AA$, and the measured inverse
momentum widths of the L=1 SL peaks, equal to $\rm 525 \pm 50~\AA$
and $\rm 620 \pm 50~\AA$ at the Ni and Cu edges respectively, are
consistent with the total SL thickness of $\rm 575 \pm 20~\AA$. In
addition, self-absorption has a clear signature in FY in the form of
an increase in the intensity above the edge~\cite{1993Eisebitt}, not
seen in measurements on the LNO-LCO SL at the Cu edge (Fig. 6c). The
slope in FY in Fig. 7c (inset) is due to a peak in the undulator
spectrum. The relatively small self-absorption and the RSXS
scattering geometry away from grazing incidence (the detector angles
were $19.8^{\circ}$ and $37.6^{\circ}$ for L=1 and L=2 reflections
at the Ni edge) support the use of a single-scattering model for the
analysis of the RSXS measurements.

The contribution to scattering from, for instance, far-away
resonances and the relatively small difference between the $c$-axis
parameters of LNO and LCO is negligible because it would give a
non-resonant energy intensity tail at $L$=2 near Ni edge, which is
not observed (Fig. 6a). $S_{2}$, the $L$=2 structure factor in Eq.
1, will then be proportional to $\Delta=|\sum_{l} p^{3+}_l e^{i 4\pi
z_{l}/c}|$, the $L$=2 component of the Fourier transform of the set
$\{p^{3+}_l\}$.

In addition, $S_{2}$ will be proportional to $f_{(2+) \rightarrow
(3+)}$, giving the energy variation of $L=2$ scattering. However,
the $L$=2 scattering is clearly made of only one peak (Figs. 7c and
8a) and the relatively small scattering angles suppress the
scattering intensity from the doped $d_{x^{2}-y^2},\downarrow$
states, which suggests that we can neglect this scattering channel.
The contributions to scattering from terms proportional to $f^{xx}$
and $f^{yy}$ are suppressed at small scattering angles and the
contribution from a small $\sigma$-polarized beam component is
neglected. Then, the only terms left from Eqs. 2 and 3 are those
proportional to $f^{zz}_{d_{3z^{2}-r^{2}},\uparrow}$ and
$f^{zz}_{d_{3z^{2}-r^{2}},\downarrow}$.

\begin{figure}
\centering\rotatebox{0}{\includegraphics[scale=0.65]{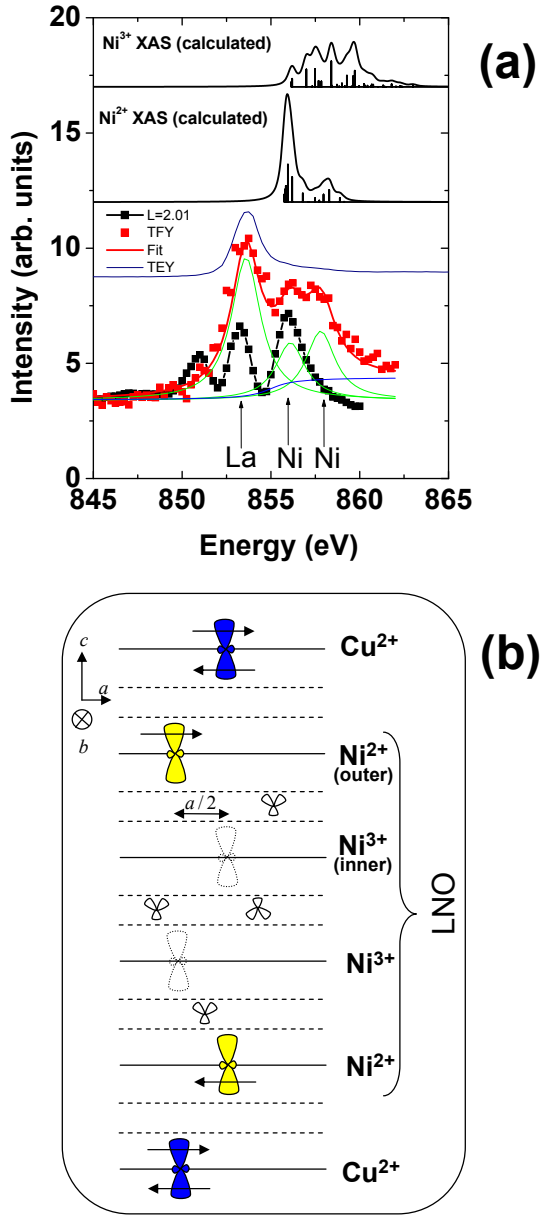}}
\caption{\label{fig:Figure8} (color online) (a) Calculated XAS
(offset vertically for clarity) using the software
``Missing"~\cite{ESRF Missing} for the $\rm Ni^{3+}$ and $\rm
Ni^{2+}$ valences are in good agreement with
Ref.~\cite{1990deGroot}. The parameters used were $10Dq=1~\rm eV$,
$Ds=Dt=0$, 0.8 Slater integrals scaling, 0.2 eV Lorentzian and 0.1
eV Gaussian broadenings. Instead of the full tetragonal $D_{4h}$
symmetry, which further splits the final $d$-states, a cubic crystal
field, with $d$-states of octahedral $O_h$ symmetry, has been used
for simplicity. The lower panel shows the measured electron yield
(navy line), fluorescence yield (red squares) at La $\rm M_4$ and Ni
$\rm L_3$ edges, and the fit of TFY (red line) with three peaks at
the La and Ni edges (green) on a smooth step function (blue). The
scattering at $L$=2 (black squares) peaks at the first Ni peak only.
The lowest energy peak at $\rm 851~eV$ is a fringe effect, seen more
clearly in Fig. 6a. (b) Sketch of the layer-resolved orbital
occupation where, for clarity, a LNO layer with 4 ML has been
illustrated and only the $d_{3z^{2}-r^{2}}$ orbitals have been
included. Dashed lines represent the double LaO layers between which
oxygen interstitials are located. Their number has been exaggerated
in this sketch and the tri-lobe drawing (not to scale) illustrates
their hybridization with La ions in the two adjacent LaO layers
along axes at $90^{\circ}$ with each other.~\cite{1989Jorgensen} A
small spin gradient is also shown at the interface.}
\end{figure}

The discussion in Sec. III A is independent of the specific details
of the SL structure, given by the $\{s_{L}\}$ distribution. However,
to obtain a numerical estimate, it is necessary to refer to the SL
structure from our HXR measurements. The intensities of $L$=2 and
$L$=1 scattering for an 8 ML superlattice at the Ni edge given by
Eq. 1 are:
\begin{align}
I(L=2)|_{Ni}=A_{L=2}(\Delta)^{2}\bigg|\frac{f^{zz}_{d_{3z^{2}-r^{2}},\downarrow}}{2}\bigg|^{2}\rm
cos^{4}(\theta_{L=2})\label{eqn:Equation6}
\end{align}
\begin{align}
I(L=1)|_{Ni}=A_{L=1}(2.61)^{2}|f^{zz}_{d_{3z^{2}-r^{2}},\uparrow}|^{2}
\rm cos^{4}(\theta_{L=1})\label{eqn:Equation7}
\end{align}
\noindent where $|\sum_{l=1}^{8}s_{l}e^{i2\pi z_{l}
/c}|=|\sum_{l=1}^{4}e^{i2\pi z_{l} /c}|=2.61$ and the smaller
contribution to $S_1$ proportional to $\Delta$ has been neglected.
Using $f^{zz}_{d_{3z^{2}-r^{2}},\downarrow} \approx
f^{zz}_{d_{3z^{2}-r^{2}},\uparrow}$ (from, for instance, the small
XMLD for $\rm Ni^{2+}$ ions~\cite{2007Arenholz}) and neglecting
footprint effects ($A_{L=1}\approx A_{L=2}$), the ratio of $L$=2 to
$L$=1 scattering intensity is:
\begin{eqnarray}
\frac{I(L=2)}{I(L=1)}\bigg|_{Ni}=\frac{(\frac{\Delta}{2})^{2} \rm
cos^{4}(\theta_{L=2})}{(2.61)^{2} \rm
cos^{4}(\theta_{L=1})}.\label{eqn:Equation8}
\end{eqnarray}
The scattering at L=2 is well separated in energy at the Ni and La
edges (Fig. 6). However, this is not the case for
L=1.~\cite{SUPPL-FIG} In practice, to obtain the value of the
scattering intensity for L=1 at the Ni edge, we cut through the
2-dimensional profile at the energy where the L=2 scattering has the
maximum intensity. This is justified by the simple shape of the SL
resonance profiles at the Ni edge. The small contribution from the
La edge tail is neglected. From the measurements
$\frac{I(L=2)}{I(L=1)}\Big|_{Ni}=0.0185$, $\theta_{L=1}=9.9^{\circ}$
and $\theta_{L=2}=18.8^{\circ}$. Then, from Eq. 7, $\Delta=0.77$, or
a change in the average Ni ion valence between the inner and outer
layers of $(p^{3+}_{in}-p^{3+}_{out})=\Delta/\sqrt{2}=0.54$.

The calculations above were made for an ideal superlattice, with no
interfacial roughness. STEM~\cite{2002Ohtomo,2006Hamann} and HXR
measurements (e.g., in Ref. 5) have shown that interface mixing in
complex oxide layers is well approximated by a Gaussian roughness.
STEM measurements on the LNO-LCO SL (data not shown) are consistent
with this assumption. We now consider a Gaussian intermixing
distribution at the interfaces. Using as upper and lower limits the
surface roughness $\sigma=\rm 0.53~nm$ (Section II.A) and $1/2$ ML,
because of the mixing of 7 ML and 8 ML superperiods, we obtain an
interface roughness $\sigma_{i}=4.25 \pm 1~\rm \AA$. Including
roughness effects in a scattering model suppresses the scattering
intensity at $L=2$ more than at $L=1$ and increases the value for
$(p^{3+}_{in}-p^{3+}_{out})$ calculated from the measurements.
Specifically, $\Delta_{R}$ (with roughness included) is $\rm
\Delta_{R}=\sqrt{e^{3(2\pi \sigma_{i}/c)^2}}\Delta \approx 1.49
\Delta$ and $(p^{3+}_{in}-p^{3+}_{out})_{\rm{R}}=0.8$.

HXR measurements showed that the superlattice is made of both 8 ML
and 7 ML superperiods (Section II.A); a similar calculation as the
above for a 7 ML superperiod gives
$(p^{3+}_{in}-p^{3+}_{out})_{\rm{R}}=1.1$. The average of the two
structures is $0.95$. These calculations show a valence change of
$50-100~\%$, pointing to a considerable modulation of the Ni valence
between the inner and the outer LNO layers, much larger than that
expected from chemical bonding of $1-10~\%$. The large valence
change is possible in our case because of the high doping of the LNO
layers with interstitial oxygen. Choosing a larger valence for Ni
ions inside the LNO layers is justified by the oxygen affinity of
LNO; the Ni $d_{3z^2-r^2}$ and $d_{x^2-y^2}$ orbitals are emptied
preferentially away from the interfaces. The case of a valence
change of $1$ is represented schematically in Fig. 8b.

The layer scattering factor was obtained above by interpolating
between nominal $\rm Ni^{2+}$ and $\rm Ni^{3+}$ form factors
statistically with weights $\{ p_{l} \}$. The same results obtain
for a scattering model describing a fractional valence common to all
Ni ions in a $\rm NiO_{2}$ layer, in which a linear dependence of
the Ni ion form factor on doping for intermediate valences between
$\rm Ni^{2+}$ and $\rm Ni^{3+}$ is used. Which model better
represents the reality depends on whether the oxygen interstitials
dope only the immediate Ni neighbor ions or a large number of ions
in the $\rm NiO_2$ layers. Our experiment cannot distinguish between
these two cases.

The large Ni valence modulation would suggest a similar effect in
the valence of the oxygen ions, in addition to their layer-dependent
density shown in Fig. 8b. One possible indication of this effect is
the feature seen at 533.5 eV (Fig. 5). It is a state in the LNO
layers since it aligns with a peak in TEY spectrum (Fig. 4). In
contrast to the LNO MCP state, it has a large intensity at $L=2$,
similar to the Ni edge scattering. It has also been observed for
very high doping in the $\rm La_{2-x}Sr_{x}CuO_{4}$ system, which
has no interstitial oxygen.~\cite{S2011B} This suggests that it is
an apical oxygen state hybridized with Ni orbitals, responding to
the large Ni valence modulation.

\section{Conclusion}

We observed a non-uniform distribution of holes in a LNO-LCO
superlattice, with doped LNO layers and essentially undoped LCO
layers. As for bulk LCO and LNO, the holes in the LCO layers are
oriented in the $a-b$ plane, while the holes in the LNO layers have
a substantial $c$-axis component. The difference in the hole
affinities of LCO and LNO layers modulates the valence and the
orbital occupation of the Ni ions in the LNO layers, with the Ni
$d_{x^{2}-y^{2},\downarrow}$ and $d_{3z^{2}-r^{2},\downarrow}$
orbital states preferentially occupied near the interfaces. Since
the density of oxygen interstitials can be continuously modified
with standard techniques, the Ni orbital occupation near the
interfaces in 214 LNO-LCO superlattices is tunable. In contrast to
the ferromagnetism of this structure, this orbital occupation is
more robust, being stable up to at least 280 K.

It would be useful to explore combinations of 214 oxides
(manganites, titanates, etc.) in which the orbital occupations near
the interfaces can be changed by selective oxygen doping so as to
leave only one orbital unoccupied, similar to superconducting
cuprates. For superconducting samples it might be desirable to avoid
ferromagnetism. On the other hand, it would be interesting to
investigate 214 systems with magnetic orders with uncompensated
spins at interfaces, for instance A-type antiferromagnets, for which
the magnetic interactions across interfaces are sufficiently strong
to correlate the spins and create a magnetically active
interface.~\cite{2005Freeland,1991Coffey}

\section{Acknowledgments}

We thank A. Gozar for useful discussions. This work was supported by
the Department of Energy: RSXS measurements by grant
DE-FG02-06ER46285, superlattice growth by MA-509-MACA, National
Synchrotron Light Source by DE-AC02-98CH10886, Materials Research
Laboratory facilities by DE-FG02-07ER46453 and DE-FG02-07ER46471,
and Advanced Photon Source by DE-AC02-06CH11357.

\end{document}